\documentclass[onecolumn,fleqn]{revtex4}

\usepackage{ifthen}
\usepackage{ifpdf}

\usepackage{latexsym}
\usepackage{amsmath} 
\usepackage{amssymb} 
\usepackage{bm}

\ifpdf
\usepackage{graphicx}
\usepackage{epstopdf}

\else
\usepackage{graphicx}
\usepackage{epsfig}

\fi

\usepackage[colorlinks,linkcolor=black,bookmarks=true]{hyperref}

\newcommand{\trc}{\mbox{trace}}

\newcommand{\tbox}[1]{\mbox{\tiny #1}}

\newcommand{\beq}{\begin{eqnarray}}
\newcommand{\eeq}{\end{eqnarray}} 

\newcommand{\hide}[1]{}

\newcommand{\addsubsectiontocontents}[1]
{\phantomsection \addcontentsline{toc}{subsection}{#1}}

\renewcommand{\thesection}{\arabic{section}}
\renewcommand{\thesubsection}{\arabic{subsection}}
\newcommand{\sheadB}[1]
{
\addtocounter{section}{1}
\setcounter{subsection}{0} 
\addsubsectiontocontents{\thesection \ \ #1} 
{\bf\LARGE[\thesection] \ #1}
}

\newcommand{\sheadC}[1]
{
\addtocounter{subsection}{1}
\ \\ {\Large\bf $=\!=\!=\!=\!=\!=\;$ [\thesection.\thesubsection] \ #1}
}

\begin{document} 
 
\title{EPR, Bell, Schrodinger's cat, and the Monty Hall Paradox} 
 
\author{Doron Cohen} 
 
\affiliation{Department of Physics, Ben-Gurion University, Beer-Sheva 84105, Israel} 
 
\begin{abstract} 
The purpose of this manuscript is to provide a short pedagogical 
explanation why ``quantum collapse" is not a metaphysical event, 
by pointing out the analogy with a ``classical collapse" which 
is associated with the Monty Hall Paradox.
\end{abstract}

\maketitle 


This manuscript constitutes a short self-contained version 
of some selected sections taken from ``lecture notes in Quantum Mechanics" 
[quant-ph/0605180](${\sim}250p$).
In particular section ${47.4}$ regarding the notion of 
collapse has attracted some attention. In this section I suggest     
to use the ``Monty Hall Paradox" as a pedagogical introduction  
to the discussion of ``quantum collapse". From my experience it is 
the most effective way to convince students and other non-experts  
that ``quantum collapse" is not a metaphysical event.

\ \\ \ \\ \ \\ 
\setcounter{section}{7} 
\sheadB{Quantum States}

\sheadC{Is the world classical? (EPR, Bell)}

We would like to examine whether the world we live 
in is ``classical" or not. The notion of 
classical world includes mainly two ingredients: 
(i) realism (ii) determinism.  
By realism we means that any quantity that can 
be measured is well defined even if we do not 
measure it in practice. By determinism we mean 
that the result of a measurement is determined 
in a definite way by the state of the system 
and by the measurement setup. We shall see later 
that quantum mechanics is not classical in both 
respects: In the case of spin~$1/2$ we cannot 
associate a definite value of $\hat{\sigma}_y$ 
for a spin which has been polarized in the $\hat{\sigma}_x$  
direction. Moreover, if we measure the $\hat{\sigma}_y$
of a $\hat{\sigma}_x$ polarized spin, 
we get with equal probability $\pm1$ as the result.
 
In this section we would like to assume that our 
world is "classical". Also we would like to assume    
that interactions cannot travel faster than light.  
In some textbooks the latter is called "locality 
of the interactions" or "causality". It has been found 
by Bell that the two assumptions lead to an inequality 
that can be tested experimentally. It turns out 
from actual experiments that Bell's inequality are violated. 
This means that our world is either non-classical or else we have 
to assume that interactions can travel faster than light.

If the world is classical it follows  
that for any set of initial conditions 
a given measurement would yield a definite result. 
Whether or not we know how to predict or calculate 
the outcome of a possible measurement is not assumed.
To be specific let us consider a particle of zero spin, 
which disintegrates into two particles going
in opposite directions, each with spin~$1/2$. 
Let us assume that each spin is described 
by a set of state variables. 
\beq
\mbox{state of particle A} &=&  x^A_1, x^A_2, ... 
\\ \nonumber
\mbox{state of particle B} &=&  x^B_1, x^B_2, ...
\eeq
The number of state variables might be very big,
but it is assumed to be a finite set. Possibly we 
are not aware or not able to measure some of these ``hidden'' variables.

Since we possibly do not have total control over the
disintegration, the emerging state of the two particles 
is described  by a joint probability function 
$\rho\left(x^A_1,...,x^B_1,...\right)$.
We assume that the particles do not affect each other  
after the disintegration (``causality" assumption).
We measure the spin of each of the particles using
a Stern-Gerlach apparatus. The measurement can yield
either $1$ or $-1$. For the first particle the measurement 
outcome will be denoted as $a$, and for the second particle 
it will be denoted as $b$.  It is assumed that 
the outcomes $a$ and $b$ are determined in a deterministic 
fashion. Namely, given the state variables of the particle 
and the orientation $\theta$ of the apparatus we have 
\beq
a &=& a(\theta_A) = f(\theta_A, x^A_1, x^A_2, ... ) = \pm1 
\\ \nonumber
b &=& b(\theta_B) = f(\theta_B, x^B_1, x^B_2, ... ) = \pm1
\eeq
where the function $f()$ is possibly very complicated. 
If we put the Stern-Gerlach machine in a different orientation 
then we will get different results:
\beq
a' &=&  a(\theta_A') = f\left(\theta_A', x^A_1, x^A_2, ...  \right)=\pm1 
\\ \nonumber
b' &=&  b(\theta_B') = f\left(\theta_B', x^B_1, x^B_2, ...  \right)=\pm1
\eeq
We have following innocent identity:
\beq
ab+ab'+a'b-a'b' = \pm 2
\eeq
The proof is as follows: if $b=b'$ the sum is $\pm 2a$, 
while if $b=-b'$ the sum is $\pm 2a'$. 
Though this identity looks innocent, it is completely 
non trivial. It assumes both "reality" and "causality"  
This becomes more manifest if we write this identity as 
\beq
 a(\theta_A)  b(\theta_B)
+a(\theta_A)  b(\theta_B')
+a(\theta_A') b(\theta_B)
-a(\theta_A') b(\theta_B') = \pm 2
\eeq
The realism is reflected by the assumption 
that both $a(\theta_A)$ and $a(\theta_A')$  
have definite values, though it is clear 
that in practice we can measure 
either $a(\theta_A)$ or $a(\theta_A')$, but not both.
The causality is reflected by assuming 
that $a$ depends on $\theta_A$ but not 
on the distant setup parameter $\theta_B$.

Let us assume that we have conducted this experiment many times. 
Since we have a joint probability distribution $\rho$, 
we can calculate average values, for instance:
\beq
\langle ab\rangle = \int{
\rho\left(x^A_1,...,x^B_1,...\right)
f\left(\theta_A, x^A_1,...\right)
f\left(\theta_B, x^B_1,...\right)}
\eeq
Thus we get that the following inequality should hold: 
\beq
\left|
\langle ab\rangle +\langle ab'\rangle +\langle a'b\rangle -\langle a'b'\rangle 
\right|
\leq2\eeq
This is called Bell's inequality.  
Let us see whether it is consistent 
with quantum mechanics. We assume that all the pairs 
are generated in a singlet (zero angular momentum) state.  
It is not difficult to calculate the expectation values. 
The result is   
\beq
\langle ab \rangle 
\ \ = \ \ -\cos(\theta_A-\theta_B) 
\ \ \equiv \ \ C(\theta_A-\theta_B)
\eeq
we have for example
\beq
C(0^o) = -1, 
\ \ \ \ \ \ \ \ \ \
C(45^o) = -\frac{1}{\sqrt{2}}, 
\ \ \ \ \ \ \ \ \ \
C(90^o) = 0, 
\ \ \ \ \ \ \ \ \ \
C(180^o) = +1. 
\eeq
If the world were classical the Bell's inequality would imply  
\beq
|C(\theta_A-\theta_B)+C(\theta_A-\theta_B')
+C(\theta_A'-\theta_B)+C(\theta_A'-\theta_B')| \le 2
\eeq
Let us  take $\theta_A=0^o$ 
and $\theta_B=45^o$ and  $\theta_A'=90^o$ and $\theta_B'=-45^o$. 
Assuming that quantum mechanics holds we get  
\beq
\left|
 \left(-\frac{1}{\sqrt{2}}\right)
+\left(-\frac{1}{\sqrt{2}}\right)
+\left(-\frac{1}{\sqrt{2}}\right)
-\left(+\frac{1}{\sqrt{2}}\right)
\right|
\ = \ 2\sqrt{2} \ > \ 2
\eeq
It turns out, on the basis of celebrated experiments 
that Nature has chosen to violate Bell's inequality. 
Furthermore it seems that the results of the experiments 
are consistent with the predictions of quantum mechanics.
Assuming that we do not want to admit that interactions 
can travel faster than light it follows that our world   
is not classical.

\sheadC{The four Postulates of Quantum Mechanics}

The 18th century version classical mechanics can be derived 
from three postulates: The three laws of Newton.
The better formulated 19th century version 
of classical mechanics can be derived from three postulates: 
(1) The state of classical particles is determined 
by the specification of their positions and its velocities; 
(2) The trajectories are determined by a minimum 
action principle. (3) The form of the Lagrangian 
of the theory is determined by symmetry considerations,  
namely Galilei invariance in the non-relativistic case.
See the Classical Mechanics book of Landau and Lifshitz for details.

Quantum mechanically requires four postulates: 
Two postulates define the notion of quantum state, 
while the other two postulates, in analogy with 
classical mechanics, are about the laws that 
govern the evolution of quantum mechanical systems.
[The rest of this section can be found in the lecture notes].

\sheadC{What is a Pure State} 

"Pure states" are states that have been filtered. 
The filtering is called "preparation". 
For example: we take a beam of electrons. 
Without "filtering" the beam is not polarized. 
If we measure the spin we will 
find (in any orientation of the measurement apparatus) 
that the polarization is zero. On the other hand, 
if we "filter" the beam (e.g. in the left direction) 
then there is a direction for which we will get 
a definite result (in the above example, in the right/left direction). 
In that case we say that there is full polarization - a pure state. 
The "uncertainty principle" tells us that if in a specific 
measurement we get a definite result 
(in the above example, in the right/left direction), 
then there are different measurements 
(in the above example, in the up/down direction) 
for which the result is uncertain. 
The uncertainty principle is implied by the first postulate.

\sheadC{What is a Measurement} 

In contrast with classical mechanics, in quantum mechanics measurement 
only has meaning in a statistical sense. We measure "states" 
in the following way: we prepare a collection of systems that 
were all prepared in the same way. We make the measurement 
on all the "copies". The outcome of the measurement is an 
event ${\hat{x} = x}$ that can be characterized by a distribution 
function. The single event has no statistical meaning. 
For example, if we measured the spin of a single electron and 
get ${\hat{\sigma_z} = 1}$, it does not mean that the state is polarized "up". 
In order to know if the electron is polarized we must measure 
a large number of electrons that were prepared in an identical way. 
If only $50\%$ of the events give ${\hat{\sigma_z} = 1}$ we should  
conclude that there is no definite polarization in the direction we measured!

\sheadC{Random Variables} 

A random variable is an object that can have 
any numerical value. In other words ${\hat{x} = {x}}$
is an event. Let's assume, for example, that 
we have a particle that can be in one of five 
sites: ${x = 1 , 2 , 3 , 4 , 5}$. An experimentalist 
could measure ${\mbox{Prob}( \hat{x} = 3 )}$ 
or ${\mbox{Prob}( \hat{p} = 3(2\pi/5) )}$. 
Another example is a measurement of the 
probability ${\mbox{Prob}( \hat{\sigma_z} = 1)}$ 
that the particle will have spin up. 

The collection of values of ${x}$ is called the spectrum 
of values of the random variable. We make the distinction 
between random variables with a discrete spectrum, 
and random variables with a continuous spectrum. 
[The rest of this section can be found in the lecture notes].

\sheadC{Quantum Versus Statistical Mechanics}

Quantum mechanics stands opposite classical statistical mechanics. 
A particle is described in classical statistical mechanics by a probability 
function (for presentation purpose we treat $\hat{x}$ and $\hat{p}$ 
as having discrete spectrum):
\beq \label{e1}
\rho(x,p) \ \ = \ \ \mbox{Prob}\{\hat{x}=x, \,\hat{p}= p\} 
\eeq
The expectation value of a random 
variable ${\hat{A} = A(\hat{x}, \hat{p})}$ 
is calculated using the definition:
\beq \label{e2}
&& \langle \hat{A} \rangle 
\ \ = \ \ \sum_{x,p} \rho(x, p) A(x,p)  
\ \ \equiv \ \ \trc(\rho A) 
\\
\label{e3}
\mbox{In particular we can write:} 
\ \ \ \ \ \ \ \ \ \ &&
\rho(x,p) \ \ = \ \ \langle \ \delta(\hat{p}-p) \ \delta(\hat{x}-x) \ \rangle
\eeq
From the definition of the expectation value follows 
the linear relation ${\langle \alpha \hat{A} + \beta \hat{B} \rangle 
= \alpha \langle \hat{A}\rangle + \beta \langle \hat{B} \rangle}$
for any pair of observables.
This linear relation is a trivial result of classical probability 
theory. {\em It assumes that the joint probability function 
Eq.(\ref{e1}) can be defined}. But in quantum mechanics we 
cannot define a ``quantum state" using a joint 
probability function, as implied by the observation  
that our world is not ``classical". For this reason, 
we have to use a more sophisticated approach. 
Loosely speaking one may say that Quantum~Mechanics 
takes Eq.(\ref{e3}) as {\em the} definition of~${\rho}$, 
and use the linear relation of the expectation 
values as a {\em second postulate} in order to deduce Eq.(\ref{e2}).

[The rest of this section an be found in the lecture notes].

\ \\ \ \\ \ \\ 
\setcounter{section}{46}
\sheadB{Theory of Quantum Measurements}
\setcounter{subsection}{3}

\sheadC{Measurements, the notion of collapse}

In elementary textbooks the quantum measurement process 
is described as inducing ``collapse" of the wavefunction. 
Assume that the system is prepared in state 
${\rho_{\tbox{initial}}=|\psi\rangle \langle \psi|}$ and that one 
measures ${\hat{P}=|\varphi\rangle \langle \varphi|}$. If the result 
of the measurement is $\hat{P}=1$ then it is said 
that the system has collapsed into the 
state  ${\rho_{\tbox{final}}=|\varphi\rangle \langle \varphi|}$. 
The probability for this ``collapse" is given by the projection 
formula ${\mbox{Prob}(\varphi | \psi) = |\langle \varphi | \psi \rangle|^2}$.

If one regard $\rho(x,x')$ or $\psi(x)$ as representing {\em physical reality},  
rather than a probability matrix or a probability amplitude, 
then one immediately gets into puzzles. Recalling the EPR experiment 
this world imply that once the state of one spin is measured at Earth, 
then immediately the state of the other spin (at the Moon) 
would change from unpolarized to polarized.  This would suggest that 
some spooky type of ``interaction" over distance has occurred. 

In fact we shall see that the quantum theory of measurement 
does not involve any assumption of spooky ``collapse" mechanism. 
Once we recall that the notion of quantum state has 
a statistical interpretation the mystery fades away. 
In fact we explain (see below) that {\em there is ``collapse" also in classical physics}!
To avoid potential miss-understanding it should be clear 
that I do not claim that the classical ``collapse" 
which is described below is an explanation of the 
the quantum collapse. The explanation of quantum collapse  
using a quantum measurement (probabilistic) point of view will 
be presented in a later section. The only claim of this section 
is that in probability theory a correlation is frequently mistaken 
to be a causal relation: ``smokers are less likely to have Alzheimer" not because cigarettes 
help to their health, but simply because their life span is smaller. 
Similarly quantum collapse is frequently mistaken to 
be a spooky interaction between well separated systems.

Consider the thought experiment 
which is known as the ``Monty Hall Paradox".
There is a car behind one of three doors. 
The car is like a classical "particle", 
and each door is like a "site". 
The initial classical state is such that 
the car has equal probability to be behind 
any of the three doors. You are asked to make a guess.
Let us say that you peak door \#1. Now the organizer 
opens door \#2 and you see that there is no car behind it. 
This is like a measurement. Now the organizer allows 
you to change your mind. The naive reasoning is that 
now the car has equal probability to be behind either 
of the two remaining doors. So you may claim that 
it does not matter. But it turns out that this simple 
answer is very very wrong! The car is no longer 
in a state of equal probabilities: Now the probability 
to find it behind door \#3 has increased. A standard   
calculation reveals that the probability to find 
it behind door \#3 is twice large compared with 
the probability to find it behind door \#2.
So we have here an example for a classical collapse. 

If the reader is not familiar with this well known "paradox", 
the following may help to understand why we have this 
collapse (I thank my colleague Eitan Bachmat for providing 
this explanation).  Imagine that there are billion doors. 
You peak door \#1. The organizer opens all the other doors 
except door \#234123. So now you know that the 
car is either behind door \#1 or behind door \#234123. 
You want the car. What are you going to do?    
It is quite obvious that the car is almost definitely 
behind door \#234123. It is also clear the that 
the collapse of the car into site \#234123 does not 
imply any physical change in the position of the car.

\sheadC{Quantum measurements, Schroedinger's cat}

What do we mean by quantum measurement? In order to 
clarify this notion let us consider a system and a detector 
which are prepared independently as 
\beq
\Psi = \left[ \sum_a \psi_a |a\rangle \right] \otimes | q=0 \rangle 
\eeq
As a result of an interaction we assume that 
the detector correlates with the system as follows:
\beq
\hat{U}_{\tbox{measurement}}\Psi = \sum\psi_a |a\rangle  \otimes  | q=a \rangle 
\eeq
We call such type of unitary evolution "ideal measurement". 
If the system is in a definite $a$ state, then it is not 
affected by the detector. Rather, we gain information on 
the state of the system. One can think of $q$ as representing 
a memory device in which the information is stored.
This memory device can be of course the brain of a human 
observer. Form the point of view of the observer, 
the result at the end of the measurement process 
is to have a definite $a$.  This is interpreted 
as a "collapse" of the state of the system. Some people 
wrongly think that "collapse" is something that goes 
beyond unitary evolution. But in fact this term just 
makes over dramatization of the above unitary process.

The concept of measurement in quantum mechanics involves 
psychological difficulties which are best 
illustrated by considering the "Schroedinger's cat" experiment.  
This thought experiment involves a radioactive nucleus, a cat, and a human being.  
The half life time of the nucleus is an hour. 
If the radioactive nucleus decays it triggers 
a poison which kills the cat. 
The radioactive nucleus and the cat are inside 
an isolated box. At some stage the human observer 
may open the box to see what happens with the cat...   
Let us translate the story into a mathematical language.  
A time $t=0$ the state of the universe (nucleus$\otimes$cat$\otimes$observer) is 
\beq
\Psi \ = 
\ |\uparrow=\mbox{radioactive}\rangle \ \otimes 
\ |{q=1=\mbox{alive}}\rangle 
\ \otimes |{Q=0=\mbox{ignorant}}\rangle
\eeq
where $q$ is the state of the cat, and $Q$ is the state 
of the memory bit inside the human observer.    
If we wait a very long time the nucleus would 
definitely decay, and as a result we will have a 
definitely dead cat:
\beq
U_{\tbox{waiting}} \Psi \ = 
\ |\downarrow=\mbox{decayed}\rangle \ \otimes 
\ |{q=-1=\mbox{dead}}\rangle 
\ \otimes \ |{Q=0=\mbox{ignorant}}\rangle
\eeq
If the observer opens the box he/she would see a dead cat:
\beq
U_{\tbox{seeing}} U_{\tbox{waiting}} \Psi \ = 
\ |\uparrow=\mbox{decayed}\rangle \ \otimes 
\ |{q=-1=\mbox{dead}}\rangle 
\ \otimes \ |{Q=-1=\mbox{shocked}}\rangle
\eeq
But if we wait only one hour then 
\beq
U_{\tbox{waiting}} \Psi \ = 
\ \frac{1}{\sqrt{2}} \Big[ 
|\uparrow\rangle  \otimes |q=+1\rangle  + |\downarrow\rangle  \otimes |q=-1\rangle  
\Big]  
\ \otimes \ |{Q=0=\mbox{ignorant}}\rangle
\eeq
which means that from the point of view 
of the observer the system (nucleus+cat) 
is in a superposition. The cat at this 
stage is neither definitely alive 
nor definitely dead. But now the 
observer open the box and we have:
\beq
U_{\tbox{seeing}} U_{\tbox{waiting}} \Psi \ = 
\ \frac{1}{\sqrt{2}} \Big[ 
|\uparrow\rangle  \otimes |q=+1\rangle \otimes |Q=+1=\mbox{happy} \rangle
\ + \ 
|\downarrow\rangle  \otimes |q=-1\rangle \otimes |Q=-1=\mbox{shocked}\rangle
\Big]  
\eeq
We see that now, form the point of view of the observer,  
the cat is in a definite(!) state. This is regarded by 
the observer as ``collapse" of the superposition.
We have of course two possibilities: one possibility is that  
the observer sees a definitely dead cat, while the other 
possibility is that the observer sees a definitely alive cat. 
The two possibilities "exist" in parallel, which leads to 
the "many worlds" interpretation.  Equivalently one may 
say that only one of the two possible scenarios is realized 
from the point of view of the observer,   
which leads to the "relative state" concept of Everett.
Whatever terminology we use, "collapse" or "many worlds" 
or "relative state", the bottom line is that we have 
here merely a unitary evolution.

\sheadC{Measurements, formal treatment}

In this section we describe mathematically how an ideal measurement 
affects the state of the system. First of all let us write 
how the $U$ of a measurement process looks like. The formal expression is  
\beq
\hat{U}_{\tbox{measurement}}
\ \ = \ \ \sum_a \hat{P}^{(a)}\otimes \hat{D}^{(a)}
\eeq
where $\hat{P}^{(a)}=|a\rangle\langle a|$ is the projection 
operator on the state $|a\rangle$, and $\hat{D}^{(a)}$ is a 
translation operator. Assuming that the measurement device 
is prepared in a state of ignorance $|q=0\rangle$, 
the effect of $\hat{D}^{(a)}$ is to get $|q=a\rangle$. Hence 
\beq
\hat{U}\Psi
=\left[\sum_a \hat{P}^{(a)} \otimes \hat{D}^{(a)}\right] 
\left(\sum_{a'} \psi_{a'} |a'\rangle  \otimes  | q=0 \rangle \right)
=  \sum_a \psi_a |a\rangle \otimes \hat{D}^{(a)}|q=0\rangle
= \sum_a \psi_a |a\rangle \otimes |q=a\rangle
\eeq
A more appropriate way to describe the state 
of the system is using the probability matrix. 
Let us describe the above measurement process using this 
language. After "reset" the state 
of the measurement apparatus is ${\sigma^{(0)}=|q{=}0\rangle\langle q{=}0|}$.  
The system is initially in an arbitrary state $\rho$. 
The measurement process correlates that state of the 
measurement apparatus with the state of the system as follows:
\beq
\hat{U}\rho\otimes \sigma^{(0)} \hat{U}^\dagger
= \sum_{a,b} \hat{P}^{(a)} \rho \hat{P}^{(b)} 
\otimes [\hat{D}^{(a)}] \sigma^{(0)} [\hat{D}^{(b)}]^{\dagger}
= \sum_{a,b} \hat{P}^{(a)} \rho \hat{P}^{(b)} 
\otimes |q{=}a\rangle\langle q{=}b|
\eeq
Tracing out the measurement apparatus we get
\beq
\rho^{\tbox{system}}  
\ \ = \ \ 
\sum_a \hat{P}^{(a)} \rho^{\tbox{preparation}}  \hat{P}^{(a)} 
\ \ = \ \ 
\sum_a p_a \rho^{(a)}
\eeq
Where $p_a$ is the trace of the projected probability 
matrix $\hat{P}^{(a)} \rho \hat{P}^{(a)}$, while $\rho^{(a)}$ 
is its normalized version.
We see that the effect of the measurement is to turn the 
superposition into a mixture of $a$ states, 
unlike unitary evolution for which 
${
\rho^{\tbox{system}}  
= 
U_{\tbox{system}} \  \rho^{\tbox{preparation}} \ U_{\tbox{system}}^{\dag}
}$.
So indeed a measurement process looks like a non-unitary process: 
it turns a pure superposition into a mixture. 
A simple example is in order. Let us assume that the 
system is a spin 1/2 particle. The spin is prepared 
in a pure polarization state $\rho=\mid \psi \rangle \langle \psi \mid$ 
which is represented by the matrix 
\beq
\rho_{ab} = \psi_a \psi_b^{*} 
=  \left( \begin{array}{cc}
\mid \psi_1 \mid^2 & \psi_1 \psi_2^{*} \\
\psi_2 \psi_1^{*} & \mid \psi_2 \mid^2 
\end{array}\right)
\eeq
where $1$ and $2$ are (say) the "up" and "down" states.
Using a Stern-Gerlech apparatus we can measure 
the polarization of the spin in the up/down direction.
This means that the measurement apparatus projects 
the state of the spin using     
\beq
P^{(1)} = \left( \begin{array}{cc}
1 & 0 \\
0 & 0 \end{array} \right)
\hspace{15mm} \mbox{and} \hspace{15mm}
P^{(2)} = \left( \begin{array}{cc}
0 & 0 \\
0 & 1 \end{array} \right)
\eeq
leading after the measurement to the state 
\beq
\rho^{\tbox{system}} 
\ \ = \ \ 
P^{(1)} \rho^{\tbox{preparation}} P^{(1)} + P^{(2)} \rho^{\tbox{preparation}} P^{(2)} 
\ \ = \ \ 
\left( \begin{array}{cc}
\mid \psi_1 \mid^2 & 0 \\
0 & \mid \psi_2 \mid^2 \end{array} \right)
\eeq
Thus the measurement process has eliminated the off-diagonal terms
in $\rho$ and hence turned a pure state into a mixture.  
It is important to remember that this non-unitary non-coherent evolution 
arise because we look only on the state of the system. 
On a universal scale the evolution is in fact unitary.

\end{document}